# Structural simplicity and complexity of compressed calcium: electronic origin

Valentina F. Degtyareva

Institute of Solid State Physics, Russian Academy of Sciences, Chernogolovka, Moscow province, 142432, Russian Federation

Correspondence email: degtyar@issp.ac.ru

**Synopsis**  Transition of Ca under pressure from close-packed structures *fcc*, *bcc* to open simple cubic structure is provided by core electron overlap with the valence band. Ca-VII and $In_5Bi_3$ exhibits similar crystal structures due to common features of their electronic structure.

**Abstract**  Simple cubic structure with one atom in the unit cell found in compressed calcium is contrintuitive with regards to traditional view on tendency of transition to densely packed structures on the increase of pressure. To understand this unusual transformation it is necessary to assume electron transfer from outer core to the valence band and increase of valence electron number for calcium from 2 to ~3.5. This assumption is supported by the model of the Fermi sphere – Brillouin zone interaction that increases under compression. Recently found structure of Ca-VII with tetragonal cell containing 32 atoms (*tI*32) is similar to the intermetallic compound $In_5Bi_3$ with 3.75 valence electrons per atom. Structural relations are analyzed in regard to a resemblance of the electronic structure. Correlations of structure and physical properties of Ca are discussed.

## 1. Introduction

Crystal structure and properties of elements display considerable changes under high pressure as have been investigated in the range up to 200 GPa and summarised in reviews (Schwarz, 2004; McMahon & Nelmes, 2006; Degtyareva, 2006; Degtyareva, 2009). A traditional view was based on tendency to adopt under compression more dense structures with increase in packing density and coordination number (CN) that is satisfied for elements in the right part of the Periodic Table. Commonly new conception arose after discovery a series of transformations in groups I and II from close-packed structures to open, low-coordinated structures including low-symmetry types and even incommensurate structures. A representative example is observation in calcium with its ambient pressure face-centered cubic structure an unusually open simple cubic structure at pressure above 32 GPa (Olyjnik & Holzapfel, 1984). The existence of Ca-*sc* was experimentally confirmed by recent studies (Gu *et al*., 2009) with detailed analysis of transformations of Ca from the face-centered cubic (Ca-I, *fcc*) to the body-centered cubic (Ca-II, *bcc*) and to the simple cubic (Ca-III, *sc*) at 20 and 32 GPa. The *sc* structure of Ca-III is stable in the wide region of pressure up to 113 GPa and at higher pressures were found Ca-IV (*tP*8), Ca-V (*oC*8) and Ca-VI (*oP*4) with transition pressures 139 and 158 GPa (Fujihisa *et al.*, 2008; Nakamoto *et al.*, 2010). Recently the structure Ca-VII above 210 GPa was determined as a commensurate host-guest structure with the basic cell *tI*32 (Fujihisa *et al.*, 2013). Structure types are labeled here and further with their Pearson symbols except of the simplest high-symmetry structures as *fcc*, *bcc*, *hcp* and *sc*.

Structural transformations in Ca under pressure increase accompany with a significant changes of transport properties. The superconductivity was found to appear in Ca-III at 1.2 K around 50 GPa increasing with pressure up to 5 – 7 K in the region of Ca-III and to higher value of $T_c$ in the following phases (Okada *et al.*, 1996; Yabuuchi *et al.*, 2006). Peculiarity of physical properties of compressed Ca is the high temperature of superconductivity reaching 29 K at 216 GPa (Sakata *et al.*, 2011) – the highest $T_c$ onset among elements at present. Recently performed high-pressure low-temperature experiments at around 30-50 GPa (Nakamoto *et al.*, 2010; Mao *et al.*, 2010; Tse *et al.*, 2012; Li *et al.*, 2012) found the existence of two new phases at low temperature with the *Cmmm* (*oC*2) and the *I*4$_1$/*amd* (*tI*4, β-tin) structures.

Structural changes in alkaline-earth metals are attributed to the *s – d* transfer when outer *s* valence electrons hybridize with the higher empty *d* band (Skriver, 1982). As was considered by Neaton & Ashcroft (1999, 2001) under sufficient compression the overall density (and hence relative core volume) is increased and the core states come in to play which result in an increasing valence electron density in the *interstitial* regions. Thus the state of sodium will be affected by increasing core overlap at high densities, requiring to treat the 3*s*, 2*p*, and 2*s* electrons as a valence. At significant degree of compression it is necessary to consider hybridization of upper core electrons with the valence band, as was considered by Ross & McMahan (1982).

Observed in compressed Ca great variety of structures and dramatic changes of physical properties require substantial modifications of the valence band assuming hybridizations of the 4*s* valence band with the formerly empty 4*p* and 3*d* bands as well as with the upper core 3*p* electrons. As a consequence of electrons band hybridization the effective valence electron numbers increased well above the initial valence of Ca equal 2 reaching values ~3 – 3.5. This accounts for the transformation to an open and low-coordinated structure in the sequence *fcc – bcc – sc*. Assumption of higher number of valence electrons for Ca leads to understanding of stability of the *sc* phase and following complex phases within the Hume-Rothery (H-R) mechanism based on consideration of the Fermi sphere – Brillouin zone (FS-BZ) interaction suggested by Mott & Jones (1936) for binary phases in the Cu-Zn and related alloy systems. At present this FS-BZ approach is widely applied to explanation of stability for complex intermetallic structures and even of quasicrystals as summarized by Mizutani (2010).

Within the model of FS-BZ interactions the high-pressure phase of sodium with the open and complex structure $oP8$ - $Pnma$ can be understand by assumption of 2 valence electrons per atom in similarity with the binary phase AuGa (Degtyareva & Degtyareva, 2009) that is considered as the H-R phase. Similarity of structures $oC16$-$Cmca$ observed under pressure in 4 and 5 valence elements (Si, Ge and Bi) and alkali metals (K, Rb and Cs), that are monovalent at ambient pressure, allowed to assume the increase the valence number for alkali metals to satisfy the necessary H-R conditions of the stability (Degtyareva, 2013).

In this paper two high-pressure phases of calcium Ca-III (sc) and Ca-VII ($tI32$) – with the simplest and the very complex structures – are considered to attempt understanding the structures in relation to the electron transfer from the upper core level to the valence band arising out under substantial compression $V/V_o$ starting from ~0.45 for Ca-III at ~33 GPa and descending to ~0.2 for Ca-VII at 240 GPa (Gu $et\ al.$, 2009; Fujihisa $et\ al.$, 2013). Both the simplest $sc$ structure and the complex host-guest structure are analysed within the H-R mechanism considering proper Brillouin zone polyhedrons that accommodate available valence electron counts. Experimentally measured properties (resistivity, superconductivity) are analysed in relation to crystal structures.

**2. Method of analysis and theoretical background**

Discussions of crystal structure stability in elements leave aside such factors as differences of atomic size and electronegativities that are decisive for binary and multicomponent phases. The crystal structure of metallic phases is defined by two main energy contributions: electrostatic (Ewald) energy, $E_{Ew}$, and the electron band structure term, $E_{BS}$. The latter usually favours the formation of superlattices and distorted structures. The energy of valence electrons is decreased due to a formation of Brillouin planes with a wave vector $q$ near the Fermi level $k_F$ and opening of the energy pseudogap on these planes if $q_{hkl} \approx 2k_F$. Within a nearly free-electron model the Fermi sphere radius is defined as $k_F = (3\pi^2 z/V)^{1/3}$, where $z$ is the number of valence electrons per atom and $V$ is the atomic volume. This effect, known as the H-R mechanism or electron concentration rule, was applied to account for intermetallic phases in binary simple metal systems like Cu-Zn (Mott & Jones, 1936). Binary alloy phases provide a wide area to trace structural dependence on electron concentration by variation of alloy compositions when along with basic structures ($fcc$, $bcc$ and $hcp$) a series of structures arise with distortions, vacancies and superlattices (Degtyareva & Afonikova, 2013).

The number of valence electrons can be varied in elements under non-ambient conditions due to hybridization of the valence band with upper empty bands when "common assumption of strictly integer valence is not exact" (Ashcroft, 2008). This expands the area of search for structures emerging in an element under high pressures to a wide realm embracing phases in binary alloy systems. One of such findings is relation of Na-*oP*8 to AuGa (MnP-type) with 2 valence electrons per atom (Degtyareva & Degtyareva, 2009). With increase of pressure it becomes more important to meet conditions of the H-R rules following the tendency to gain in the band structure energy. Two main parts of energy, $E_{Ew}$ and $E_{BS}$, are scaled differently on the volume as $V^{1/3}$ and $V^{2/3}$, respectively, and the latter is exceeding under significant compression.

The stability of high-pressure phases in calcium are analyzed using a computer program BRIZ (Degtyareva & Smirnova, 2007) that has been developed to construct Brillouin zones or extended Brillouin-Jones zones (BZ) and to inscribe a Fermi sphere (FS) with the proper free-electron radius $k_F$. The resulting BZ polyhedron consists of numerous planes with relatively strong diffraction factor and accommodates well the FS. The volumes of BZ's and FS can be calculated within this program. The BZ filling by the electron states ($V_{FS}/V_{BZ}$) is estimated by the program, which is important for the understanding of electronic properties and stability of a given phase. For a classical H-R phase $Cu_5Zn_8$, the BZ filling by electron states is equal to 93% which is typical for many other phases stabilized by the H-R mechanism. Diffraction patterns of these phases have a group of strong reflections with their $q_{hkl}$ lying near $2k_F$ and the BZ planes corresponding to these $q_{hkl}$ form a polyhedron that is very close to the FS. The FS would intersect the BZ planes if its radius, $k_F$, is slightly larger than $q_{hkl}/2$, and the program BRIZ can visualize these intersections. One should keep in mind that in reality the FS would usually be deformed due to the BZ-FS interaction and partially involved inside the BZ. The ratio $2k_F/q_{hkl}$, called as a "truncation" factor or a "closeness" factor, is an important characteristic for a phase stabilized due to the H-R mechanism. For the H-R phases such as $Cu_5Zn_8$, this closeness factor is equal 1.015, and it can reach up to 1.05 in some other phases. Thus, with the BRIZ program one can obtain a qualitative picture and some quantitative characteristics that allow for estimations how a structure matches the criteria of the H-R rules.

Assuming that the model of FS-BZ interaction may account for the structure stability one can deduce an effective number of valence electrons for atoms in this structure. This approach may help to understand formation of unusual structures in elements under high compression when significant changes of electronic states are expected. Estimation of the valence states in a given structure from the BZ-FS configuration can be supported by analysis of the experimentally observed transport properties as is considered below in section 3.3.

## 3. High-pressure structures of Ca as Hume-Rothery phases

Two high-pressure phases of calcium are analyzed and discussed in this paper – Ca-III, *sc,* and Ca-VII, *tI*32, on the ground of they have similar structures among other materials and they can be beneficially related to H-R mechanism of structure stability. Transitions of Ca from *fcc* to *bcc* and to *sc* are accompanied by volume drops which are 3.1% and 9.8%, respectively (Gu *et al.*, 2009). There is a decrease in coordination number (CN) 12 – 8 – 6 and in packing density 0.74 – 0.68 – 0.52. Occurrence of Ca-*sc* is quite unusual on compression, it corresponds to ~13% reduction of the interatomic distance giving the value of atomic radius 1.344 Å at 33.8 GPa at compression $V/V_o$ = 0.44 (Gu *et al.*, 2009). This value of atomic radius is very close to the effective ionic radius of Ca equal 1.34 Å (for CN = 12) and 1.12 Å (for CN = 8) as estimated by Shannon (1976). For Ca-VII, *tI*32, at 241 GPa the volume compression is $V/V_o$ = 0.22 (Fujihisa *et al.*, 2013) and the shortest interatomic distance decreases to ~2.15 Å resulting in atomic radius ~1.08 Å that is much less the ionic radius 1.18 Å (for CN = 9). To understand this dramatical decrease of atomic sizes is not enough to accept only a continuous $s - d$ electron transfer and is necessary to consider participating of outer core electrons in the valence band.

**3.1. Ca-III with the simple cubic structure: electron transfer**

To estimate optimal valence electron number for the Ca-*sc* within the H-R rules we consider diffraction pattern with position $2k_F$ and construct the BZ polyhedron by planes close to $2k_F$ (Fig. 1). Assuming $z = 2$ – the ambient pressure valence of Ca – and considering planes (100) giving BZ in the form of hexahedron with the only 6 faces (Fig. 1*b*), one could not expect any satisfaction to the H-R rules. Next diffraction peak (110) gives BZ with 12 planes – the rhombic dodecahedron –, and a contact of FS with these planes occurs at ~3 valence electrons. Taking appropriate ratio $2k_F/q_{hkl}$ ~ 1.05 one could obtain an optimal valence count as $z \approx 3.5$. This polyhedron and corresponding inscribed FS are shown on Fig. 1*c*; zone filling by electron state is $V_{FS}/V_{BZ}$ = 0.875 and ratio $2k_F/q_{110}$ = 1.057 that satisfies well the H-R rules.

The only one element in the Periodic Table – polonium – crystallizes in the *sc* structure at ambient conditions and under high pressure this structure adopt group-V elements phosphorous and arsenic (see Table 1). The *sc* structure was observed in some elements and binary alloys as a metastable phase (Table 1) within the range of valence electron numbers from 2 to 6. Within this range some special domains should be selected which are optimal for the H-R rules. First, that is a domain with values 3 – 3.5 valence electrons forming while the FS is close to the BZ's planes (110) as considered above. Next domain is at $z = 4.56$ providing that the FS is in contact with edges formed by planes (110). Further domain is determined by condition $2k_F = q_{111}$ when the FS is close to (111) planes with the exact value $z = 5.44$, and assumption of the FS-BZ overlap results in $z = 6$. The latter case is realized in Po under ambient conditions. For $z = 5$ (in P and As under pressure) one can expected the formation of FS necks attracted to (111) planes.

Interestingly, the ranges of valence electrons that are optimal for stability of the *sc* structure within the H-R rules qualitatively coincide with the maxima of superconducting temperature measured experimentally for the *sc* Au-Te films after ion irradiation by Meyer & Stritzker (1979). Decisive factor in both events is high density of electron states near the Fermi level that appears when the FS is close to the Brillouin planes forming energy gaps. Thus, for the *sc* Au-Te phases first $T_c$ maximum appears at $z = 3 – 3.5$ valence electrons and next two maxima at 4.5 and ~5 – 5.3 valence electrons. These values correlate with the appearance of superconductivity in Ca at ~50 GPa and rising $T_c$ to 5 – 7 K in the region of Ca-III where 3 – 3.5 valence electrons was suggested. It should be noted that no $T_c$ maximum for *sc* Au-Te was found at $z = 2$ that supports the assumption of higher valency for Ca-*sc*. Phases with the *sc* structure ($z = 5$) of P and As under pressure become superconductors with $T_c$ reaching 13 K (at 30 GPa) and 2.4 K (at 32 GPa), respectively (Debessai *et al.*, 2008).

**3.2. Ca-VII with the commensurate host-guest structure**

Structural behaviour of Ca under compression in the post-*bcc* region differs essentially from that of heavier alkali-earth metals Sr and Ba. Only at pressure above 210 GPa Ca-VII adopts the host-guest structure (Fujihisa *et al.*, 2013) which is similar to host-guest structures found in Ba and Sr (McMahon & Nelmes, 2006). The latter two phases are incommensurate since host and guest subcells in the common tetragonal cell have irrational ratio of *c* parameters. Thus, $c_H/c_G$ was found ~1.388 for Ba-IVa (at 12 GPa) and ~1.404 for Sr-V (at 56 GPa). The Ca-VII structure may be considered as having a similar host-guest cell with a commensurate ratio 4/3. The structure solution for Ca-VII was suggested in two models as *I4/mcm*-32 and *P4$_2$/ncm*-128, the latter is a supercell of the former with dimensions 2×2×1 (Fujihisa *et al.*, 2013). Another example of a complex host-guest structure was reported recently for Ba-IVc (at 19 GPa) with $3\sqrt{2} \times 4\sqrt{2} \times 3$ supercell of the basic host-guest cell that results in 768 atoms in the unit cell (Loa *et al.*, 2012). The host-guest ratio in this structure has a commensurate value of 4/3 as in the case of Ca-VII. The basic structure for Ca-VII has tetragonal unit cell with 32 atoms, space group *I4/mcm*, and this structure has a prototype structure In$_5$Bi$_3$-*tI*32 as found in the Pauling File Database (Villars *et al.*, 2003). One more example of this structure was obtained in the In-Sb alloy phase with composition In$_5$Sb$_3$ after high pressure – high temperature treatment (7 GPa, 623 K) and quenching (Degtyareva *et al.*, 1983).

Atomic arrangement in the In$_5$Bi$_3$-*tI*32 structure is shown in Fig. 2 that is similar to the Ca-VII basic structure. The only small difference is in c/a ratios that are equal 1.484 and 1.665 respectively. For the incommensurate host-guest structures of Ba-IVa and Sr-V the unit cell is considered comprising the host subsell with 8 atoms of two antisymmetrical square-triangle nets giving $c_H$ and the guest subcell with $c_G$ equal to the atomic distance in chains. The ratio $c_H/c_G$ is rational (4/3) for In$_5$Bi$_3$, Ca-VII and Ba-IVc, but is irrational for Ba-IVa and Sr-V leading to a non-integer number of atoms in the common unit cell.

Grounding on the assumption that common crystal structure may determined by the similar electronic structure we accept for Ca in the phase $tI$32 the number of valence electrons close to that as in In$_5$B$_3$ (~3.6 – 3.75). This assumption is necessary to provide for the satisfaction to the H-R mechanism of structure stability that is analysed by the FS-BZ interaction. Diffraction pattern of Ca-$tI$32 is shown on Fig. 3 with indication of the 2k$_F$ position for $z$ = 3.6 located near reflections (314), (400), (323) and (206), the corresponding planes form a prominent polyhedron with 44 faces. Degree of BZ filling by electron states is $V_{FS}/V_{BZ}$ = 0.82 and maximal FS - BZ planes overlap is 2k$_F$/$q_{314}$ = 1.018 satisfying well to the H-R rules. It should be noted that the polyhedron formed by the first strongest reflection (213) is placed commonly inside the FS contacting with it by vertices (Fig. 3$b$) that reduces the band structure energy. It is reasonable to extend this approach to the incommensurate structures and construct BZ polyhedra for corresponding commensurate approximants. This consideration will result in necessity to assume for Ba and Sr in the host-guest structures the valency of ~ 3.6 – 3.75 electrons. Thus, for all family host-guest structures in the alkali-earth metals the assumed electron states are related to the tendency of the core – valence band overlap under significant compression.

### 3.3. Structure – properties correlations in compressed Ca

Stability of the *sc* structure in relation to the number of valence electrons in terms of the H-R rules (as was discussed in section 3.1) can be defined by closeness of the FS to BZ planes either (110) or (111) within two regions at z = 3 – 3.5 and z = 5 – 6; an additional domain is at z = 4.56 defined by the FS contact with edges of the planes (110) polyhedron. These regions of z correlate with superconducting properties found for *sc* phases in Au-Te films (Meyer & Stritzker, 1979). The superconducting behaviour of Ca-*sc* should be examined within the region of 3 – 3.5 valence electrons where the satisfaction to H-R rule was considered. An increase of T$_c$ in the region Ca-*sc* may follow two factors. It is known that conventional superconductors reveal usually a decrease of T$_c$ on compression. Appearance of superconductivity and T$_c$ increase under pressure are connected with the increase of the FS-BZ interaction and/or with the increase of the valence electron number. A remarkable example of ~10 times increase of T$_c$ was observed in the Al-Si alloy system with increase of Si solubility in the *fcc*-Al up to 20 at.% under pressure-temperature treatment (Degtyareva *et al*., 1985).

The enhancement of superconductivity is associated with an increase of resistivity. Thus, the *bcc – sc* transition in Ca is accompanied as by appearance of superconductivity as well by substantial rise of resistivity shown under static and shock compression as was discussed by Maksimov *et al*. (2005).

On further compression of Ca was found a substantial increase of $T_c$ that reached 29 K above 210 GPa in the phase Ca-VII- *tI*32 (Fujihisa *et al.*, 2013). High value of $T_c$ is related to formation of the complex phase with BZ consisted of many planes in the form of polyhedron that accommodate well the FS (Fig. 3*c*). The high degree of BZ filling by electron states ~90% corresponds to high density of electron states at the Fermi level what is responsible for superconductivity whereas a reduction of the Fermi surface area is resulted in increase of resistivity. These effects were found for the host-guest structures in Ba, Sr and similar structures in other elements as was summarized by Degtyareva (2006). Binary alloy phases with *tI*32 structure $In_5Bi_3$ and $In_5Sb_3$ are also good superconductors with $T_c$ of 4.2 K and 5.5 K, respectively (Degtyareva *et al.*, 1983).

## 4. Summary

Stability of the simple cubic structure with respect to the number of valence electrons in terms of the H-R rules is defined by closeness of the FS to the Brillouin zone planes either (110) at z = 3 – 3.5 or (111) at z = 5 – 6. The Ca-*sc* can be adjusted within the first z region whereas *sc* phases in P, As and Po are located in the second region at z = 5 and z = 6. These regions of z are characterised relatively high temperature of superconductivity. Effects of FS-BZ interaction correlate with superconducting and transport properties.

For Ca-VII at 216 GPa was shown onset of $T_c$ at 29 K (Sakata *et al.*, 2011) for the structure with the basic tetragonal cell containing 32 atoms. This structure is similar to the known binary compound $In_5Bi_3$ of *tI*32-type found also in $In_5Sb_3$. The origin of stability for this complex structure can be deduced from the H-R mechanism based on the nearly-free electron model when the FS is accommodated within the many-faced polyhedron raised from Bragg diffraction peaks of noticeable intensity. Due to energy gaps emerging on these planes the crystal energy is reduced providing the high density of electron states near the Fermi level. These effects are responsible as for phase stability as well as for the physical properties like superconductivity and resistivity.

Structural relations of Ca-VII and $In_5Bi_3$ give opportunity to imply common features of electron structures. Consideration of the core – valence band electron transfer may promote a better understanding of non-traditional behaviour of alkali and alkali-earth elements under significant compression.

**Acknowledgements**   The author thanks Dr Olga Degtyareva of the Edinburgh University for valuable discussions. Financial support through the grant "The Matter under High Energy Density" by Programme of the Russian Academy of Sciences is highly appreciated.

**Table 1** Summary of phases with the simple cubic structure observed in elements and binary alloys at different conditions

| Element / alloy [a] | Number of valence electrons | Conditions | Ref |
|---|---|---|---|
| Po | 6 | Ambient pressure | Villars et al. (2003) |
| P | 5 | High pressure (10-100GPa) | McMahon & Nelmes (2006) |
| As |  | High pressure (25-48 GPa) |  |
| Ca | 3.5 [b] | High pressure (32-113 GPa) | McMahon & Nelmes (2006) |
| Se | 6 | After shock compression | Degtyareva & Sikorov (1977) |
| Se | 6 | Thin film (sublimation in vacuum) | Andrievskii et al. (1959) |
| Metastable phases in alloys ||||
| Au – Bi (60 %Bi) [c] | 3.4 | Rapid quenching from liquid | Giessen et al. (1968) |
| Au – Bi (70-80 % Bi) | 3.8 – 4.2 |  |  |
| Au – Sb (80 %Sb) | 4.2 |  |  |
| Au – Bi (33 -100 %Bi) | 2.31 – 5 | Vapor quenching and annealing | Häussler et al. (1984) |
| Au – Sb (68-100 %Sb) | 3.73 – 5 |  |  |
| Au – Te (60-85 %Te) | 4.0 – 5.25 | Rapid cooling from the melt | Luo & Klement (1962) |
| Ag – Te (70-80 %Te) | 4.5 – 5.25 |  |  |
| Au – Te (60-85 %Te) | 4.0 – 5.25 | Rapid quenching from liquid | Tsuei & Newkirk (1969) |
| Pb – Sb (50-60 %Sb) | 4.5 – 4.6 | Rapid quenching from liquid | Borromee-Gautier et al. (1963) |
| In – Sb (60-70 %Sb) | 4.2 – 4.4 | Rapid quenching from liquid | Jordan (1963) |
| In – Sb (60 % Sb) | 4.2 | High pressure quenching | Degtyareva et al. (1983) |
| Au – Sb (73.5-84 %Sb) | 3.94 – 4.36 | Thin film, condensed | Palatnik et al. (1961) |
| Au – Te (20-90 %Te) | 2.0 – 5.5 | Thin film, ion irradiation | Meyer & Stritzker (1979) |

(a) Alloy composition is given in atomic percent. (b) Suggested in this paper. (c) Rhombohedral cell, $\alpha = 89.1°$.

**Figure 1** (*a*) Calculated diffraction pattern of Ca-III, *sc*, $Pm\bar{3}m$, P = 33.8 GPa, *a* = 2.689 Å (Gu *et al.*, 2009); positions of $2k_F$ are indicated for the number of valence electrons *z* = 2 and *z* = 3.5 by vertical lines. (*b*) The Brillouin zone of (100) planes and the inscribed FS for *z* = 2. (*c*) The Brillouin zone of (110) planes and the inscribed FS for *z* = 3.5. The BZ-FS configuration in (*c*) is satisfying the Hume-Rothery conditions (see discussion in text).

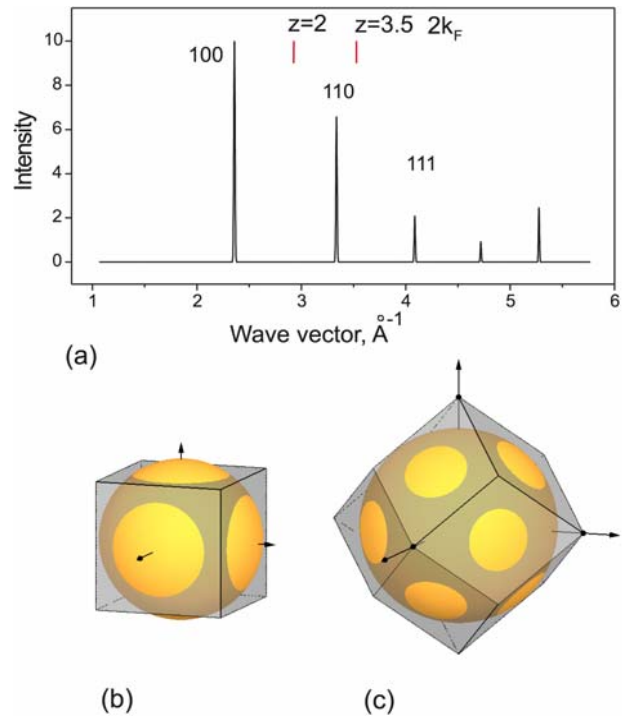

**Figure 2** (*a*) The unit cell of In$_5$Bi$_3$, *tI*32, *I4/mcm*, *a* = 8.544 Å, *c* = 12.68 Å; atoms In – blue (grey), Bi – green (dark grey). (*b*) Projection along *c*; in the *ab* base plane atoms Bi form a squares-triangle net 3$^2$434 and in the next level at height *z* ≈ 1/6 atoms In form an antisymmetricaly arranged net; these nets form antiprisms as shown in light blue (light grey). Atoms located at corners and the centre of the *ab* base plane make chains along *c* direction passing trough the centres of antiprisms in ratio when 4 distances between atoms in chains equal to 3 distances between similar layers in squares-triangle nets. The structure *tI*32 is related to the host-guest structure in Ba and Sr comprising 3 host and 4 guest subcells with the ratio $c_H/c_G$ = 4/3.

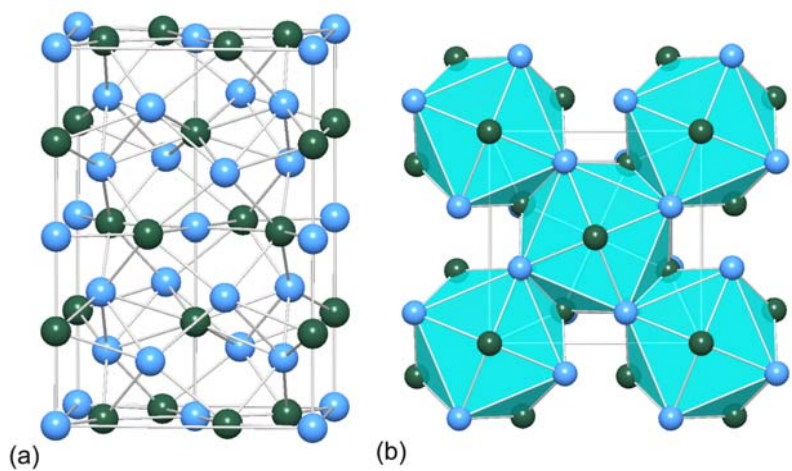

**Figure 3** (*a*) Calculated diffraction pattern of Ca-VII, *tI*32, *I4/mcm*, P = 241 GPa, *a* = 5.5099 Å, *c* = 9.1825 Å (Fujihisa *et al.*, 2013); principal diffraction peaks hkl are assigned; position of $2k_F$ is indicated for the number of valence electrons *z* = 3.6 by vertical dashed line. (*b*) The Brillouin zone of (213) planes is circumscribed by FS. (*c*) The Brillouin zone of group planes close to $2k_F$ and the inscribed FS. The BZ-FS configuration in (*c*) is satisfying the Hume-Rothery conditions (see discussion in text).

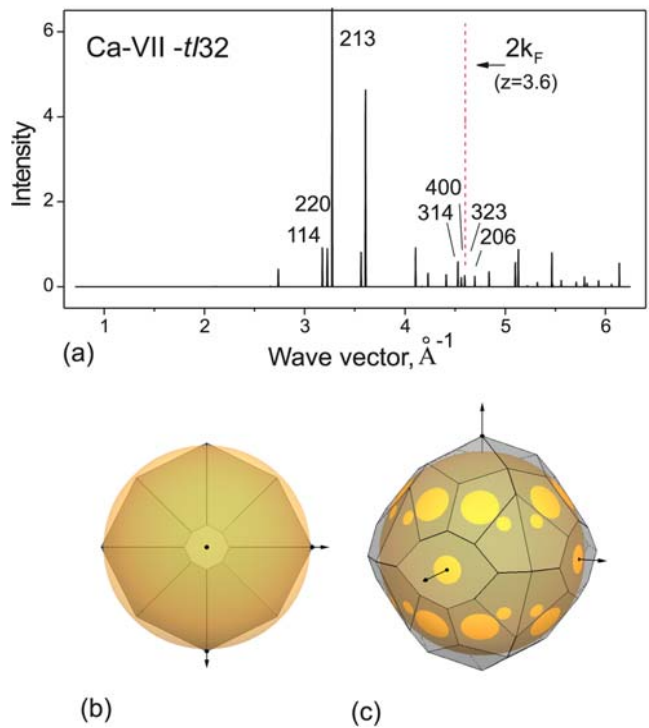